\documentclass[9pt,twocolumn,twoside]{pnas-new}

\usepackage{color}
\templatetype{pnasresearcharticle} 

\title{Burrowing dynamics of aquatic worms in soft sediments}

\author[a,1]{Arshad Kudrolli}
\author[a]{Bernny Ramirez} 

\affil[a]{Department of Physics, Clark University, Worcester, MA 01610}

\leadauthor{Kudrolli} 

\significancestatement{The physical principles of locomotion inside sediment beds at the bottom of water bodies are poorly understood because of the remoteness and opacity of the medium. We show that the common oligochaete {\textit Lumbriculus variegatus} moves faster in sediments compared with water, while using a similar combination of elongation-contraction and transverse undulatory strokes, exploiting a higher drag anisotropy of its body. Tracking the worm inside transparent sediments, we show that its speed can be captured by a linear combination of calculated speeds with resistive force theory and an anchor model of peristaltic motion.  We then demonstrate that these locomotion strategies can be observed in composting worms and can be used to move effectively in mediums with wide range of yield strengths.
}

\authorcontributions{A.K. and B.R. designed the study, obtained and analyzed the data, and wrote the results.}
\authordeclaration{Conflict of Interest: None.}
\correspondingauthor{\textsuperscript{1}To whom correspondence should be addressed. E-mail: akudrolli\@clarku.edu}


\begin{abstract} 
We investigate the dynamics of \textbf{\textit{Lumbriculus variegatus}} in water-saturated sediment beds to understand limbless locomotion in the benthic zone found at the bottom of lakes and oceans. These slender aquatic worms are observed to perform elongation-contraction and transverse undulatory strokes in both water-saturated sediments and water. Greater drag anisotropy in the sediment medium is observed to boost the burrowing speed of the worm compared to swimming in water with the same stroke using drag-assisted propulsion. We capture the observed speeds by combining the calculated forms based on resistive-force theory of undulatory motion in viscous fluids and a dynamic anchor model of peristaltic motion in the sediments.  Peristalsis is found to be effective for burrowing in non-cohesive sediments which fill in rapidly behind the moving body inside the sediment bed. Whereas, the undulatory stroke is found to be effective in water and in shallow sediment layers where anchoring is not possible to achieve peristaltic motion. We show that such dual strokes occur as well in the earthworm \textbf{\textit{Eisenia fetida}} which inhabit moist sediments that are prone to flooding.  Our analysis in terms of the rheology of the medium shows that the dual strokes are exploited by organisms to negotiate sediment beds that may be packed heterogeneously, and can be used by active intruders to move effectively from a fluid through the loose bed surface layer which fluidize easily to the well-consolidated bed below.   
\end{abstract}

\dates{This manuscript was compiled on \today}
\doi{\url{www.pnas.org/cgi/doi/10.1073/pnas.XXXXXXXXXX}}

\begin{document}

\maketitle
\thispagestyle{firststyle}
\ifthenelse{\boolean{shortarticle}}{\ifthenelse{\boolean{singlecolumn}}{\abscontentformatted}{\abscontent}}{}

Organisms burrowing in the benthic layer composed of organic and inorganic granular sediments at the bottom of water bodies can be found widely across our planet.  While the strategies used by freely swimming water-borne organisms have been well studied, those used by limbless organisms which burrow in the loose sediment bed are far less known beyond the wide use of water jets to fluidize the sediments~\cite{hosoi15}. For example, earthworms use peristalsis to move through terrestrial environments~\cite{quillin99,tanaka12},  but their use in moving through noncohesive water-saturated sediments which fluidize easily is unclear because of the difficulty in visualizing the strokes in situ.  It has been observed that undulatory body motion is employed to burrow through water-saturated sediments 
by sand lances~\cite{gidmark11} and opheliid polychaete {\it Armandia brevis}~\cite{dorgan13,dorgan15}, and it has been suggested that persistalsis may be insufficient to overcome fracture resistance or anchor small worms in unconsolidated sediments~\cite{dorgan13}. Interestingly in this context, the nematode {\it Caenorhabditis elegans} is well documented to modify its undulatory gait from high to low frequencies while swimming in water to crawling on surfaces and through agarous~\cite{fang-yen2010}. However, they appear to always undulate their bodies even while moving through loosely packed sediment monolayers~\cite{jung10}, and fixed micro-pillar arrays~\cite{lockery2008}.  

While body undulations can be readily identified across a wide range of limbless organisms~\cite{gray53}, the physical mechanism by which locomotion is accomplished varies significantly even in water depending on the size and speed of the swimmer. It is long established that the drag of the body used by the swimmer to propel itself forward can be dominated by viscous forces at low speeds and by inertia at high speeds, as measured by the Reynolds number~\cite{lighthill76,lauga09}. Measurements with spheres and rods moving in water-saturated soft sediments have found that the drag scaled by the buoyancy-subtracted weight of the grains can be used to define an effective friction $\mu_e$ which approaches a constant value $\mu_o$ at vanishing speeds, and increases many-folds with speed~\cite{panaitescu17,jewel18,allen19}. While inertia and fluid viscosity can in general both play a role in determining the rate-dependence, it has been found that inertial effects dominate in case of millimeter sized grains immersed in relatively low-viscosity fluids like water~\cite{panaitescu17,jewel18,allen19}. In this case, $\mu_e = \mu_o + k I^n$, where $k$ and $n$ are material-dependent constants, and $I$ is the inertial number. For rods~\cite{allen19}, $I = \frac{U d_g}{D \sqrt{P/\rho_g}}$, where $U$ is the body speed, $d_g$ the grain diameter, $D$ is the effective body diameter, $P$ is the pressure, and $\rho_g$ the grain density. Hence, the rheology encountered by a moving body in water-saturated sediment medium transitions from being shear rate-independent Coulomb friction-like at low speeds to rate-dependent fluid-like with increasing speed. This is different from Newtonian fluids like water, where the drag scales linearly with speed at low speeds and quadratically at higher speeds, independent of depth~\cite{balmforth14}. Thus, it is important to understand the strategies used by active intruders in water-saturated sediments in terms of the actual rheological properties of the medium. 

\begin{figure}
\begin{center}
\includegraphics[width=0.45\textwidth]{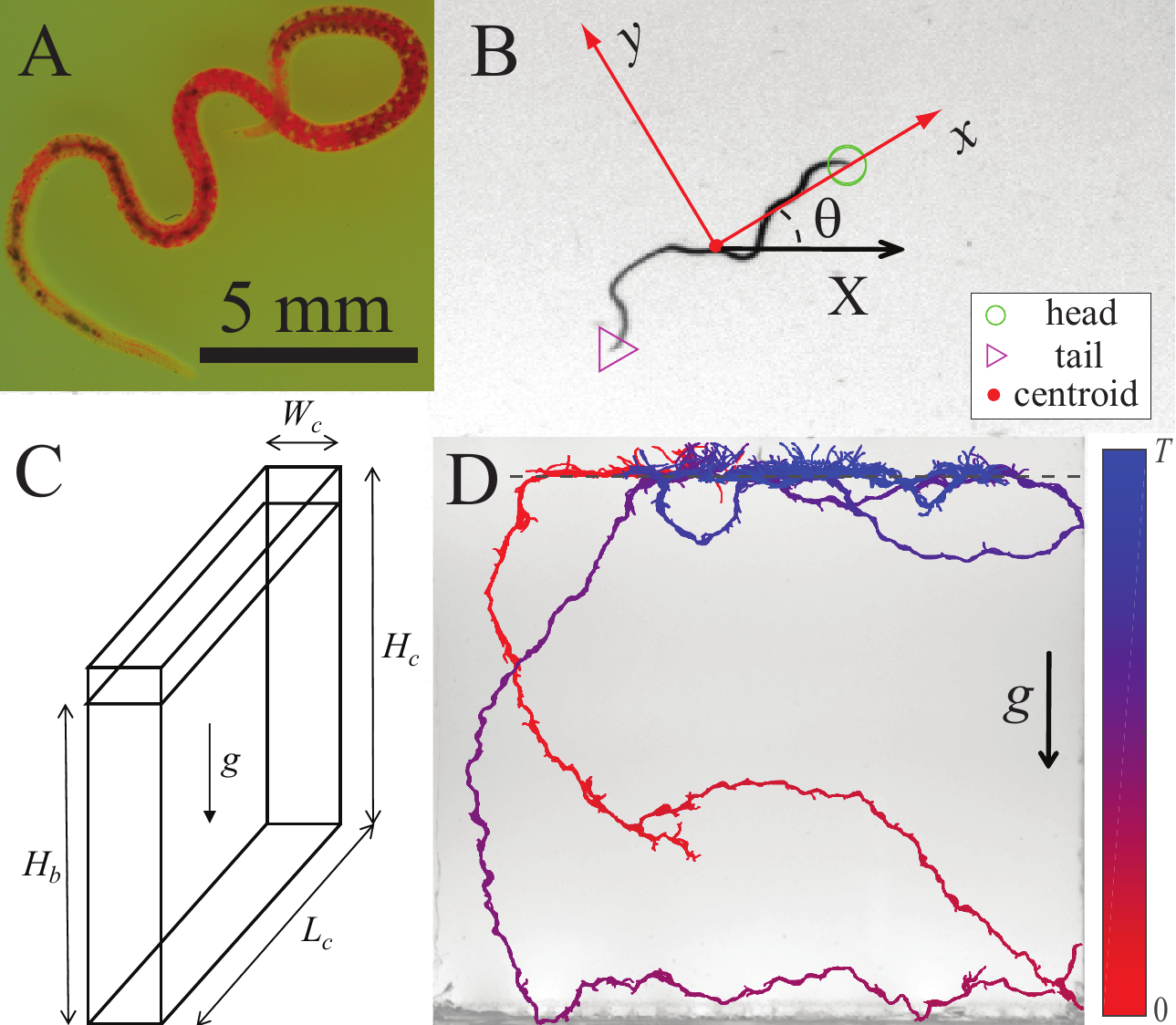}
\end{center}
\caption{(A) An Image of {\textit L. variegatus}.  (B) The tracked head, tail and centroid of a worm of length $l_w = 36.6$\,mm inside the transparent sediments. The origin in the body reference frame corresponds to the centroid, and the $x$-axis is oriented along the average angle $\theta$ that the worm subtends with the X-axis in the lab reference frame. (C) A schematic of the container filled with water and a sediment bed. (D) A sample trajectory of a worm released at the surface of the sedimented granular bed with $H_b = 20.7$\,cm, $H_c = 22.2$\,cm, $L_c = 21.5$\,cm, and $W_c = 1.27$\,cm. ($\Delta t = 1$\,s; $T = 1000$\,s). Color bar shows progress of time. The dashed line indicates the surface of the bed which is otherwise invisible.
} 
\label{fig:img}
\end{figure}

Here, we use the California blackworm {\it Lumbriculus variegatus}  (see Fig.~\ref{fig:img}A) as a paradigm to understand limbless burrowing in water-saturated soft sediments that can fluidize easily. This common freshwater oligochaete shows peristaltic motion while crawling on wet surfaces with waves of circular and longitudinal muscle contraction that move in the direction opposite to motion~\cite{drewes99a}. However, their dynamics inside noncohesive water-saturated sediments has neither been observed directly nor analyzed in terms of the drag experienced in the medium. By using clear hydrogel grains, which appear transparent when immersed inside water, we visualize the shape of the body while burrowing inside a sedimented bed (see Fig.~\ref{fig:img}B and Methods) and compare them to those used while swimming in water.  

\section*{Results}
\subsection*{Observations with \textbf{\textit{L. variegatus}}}
The projected shapes of a {\it L. variegatus} released just above a sedimented bed in a water filled container, shown schematically in Fig.~\ref{fig:img}C, is plotted at 1\,second time intervals in Fig.~\ref{fig:img}D. The worm is observed to swim above the bed surface for a few seconds before burrowing rapidly down through the bed, turning, rising, and then further exploring the surface of the sedimented bed.  Magnified views with higher time-resolution of each phase can be found in SI Appendix, Fig.~S1.  It can be observed that the worm moves in a narrow sinusoidal path while burrowing in the sediments which is not much wider than its body width. Whereas, greater lateral body motion are observed while swimming in water near the bed surface. Further, the worm can be also observed to reflect off the side walls and sometimes move backward. Similar behavior is observed as well in sedimented bed inside  thinner quasi-2D containers and wider cuboid containers (see Appendix SI, Fig.~S2). We also observe that the worms do not crawl up the side walls and slide on glass and acrylic surfaces when fully immersed in water. Thus, the container walls serve as a physical barrier, and do not appear to aid the locomotion of the worm. Because the image analysis and tracking is far simpler in 2D, we discuss worm dynamics in a container with internal dimensions $L_c = 155$\,mm, $H_c = 164$\,mm, and $W_c = 2$\,mm, which is sufficiently wide to allow the worm to move freely. The effects of the constraints imposed by lateral walls on the dynamics are further discussed in SI Appendix, Effect of Container Thickness. We focus on the locomotion dynamics when the worm is away from the side walls, and when it is essentially traveling forward without turning on itself.

\begin{figure}
\begin{center}
\includegraphics[width=0.48\textwidth]{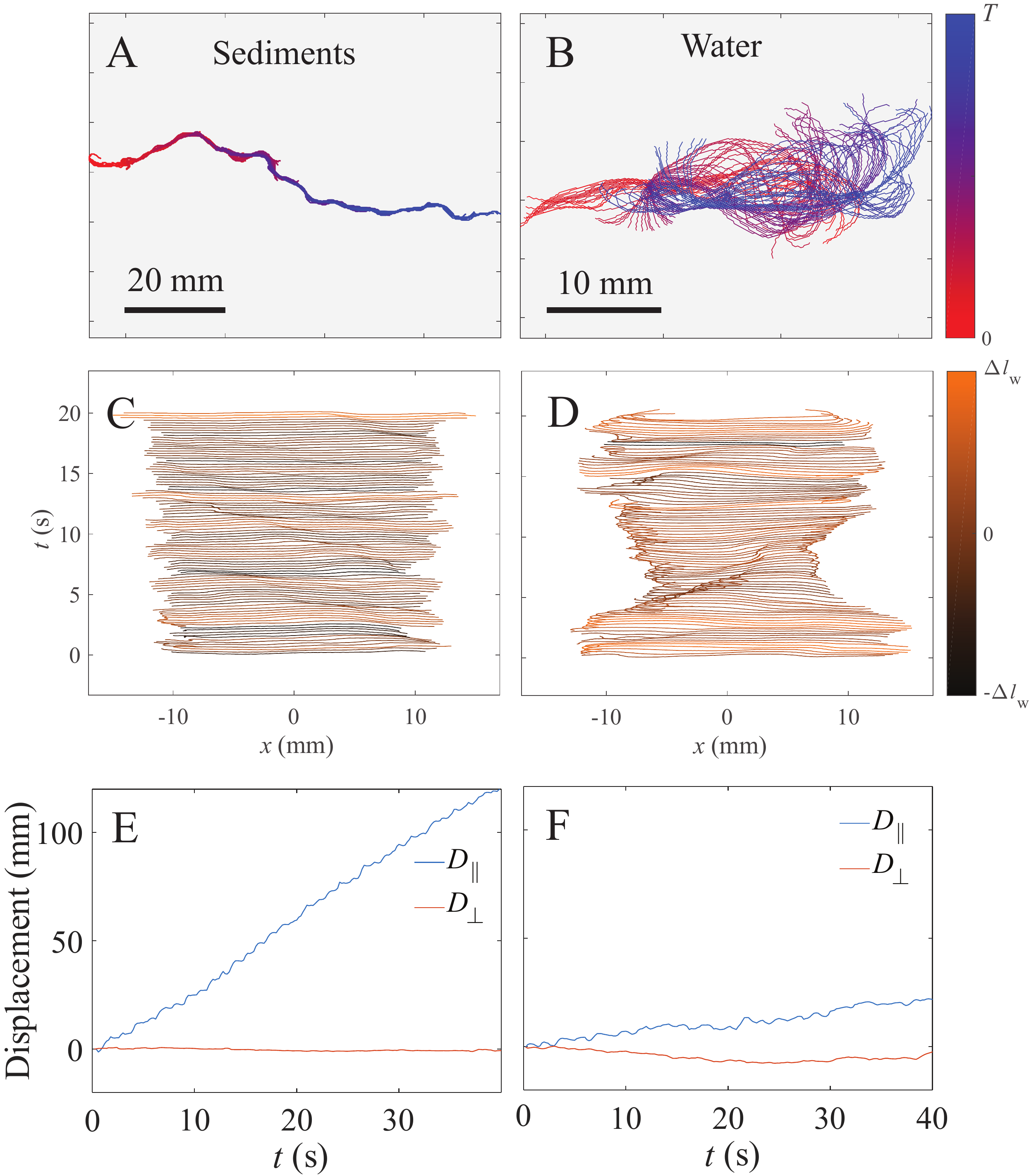}
\end{center}
\caption{(A,B) Worm trajectory in the lab frame of reference in the sediments (A), and water (B). The worm of length $l_w = 26.6$\,mm is observed to follow a narrow sinusoidal path in the sediments compared to undulating in place in water. (C,D) The corresponding worm shapes in the body frame of reference are also shown shifted up by a fixed distance over each time interval for clarity ($\Delta t =200$\,ms and $T = 20$\,s). Measured cumulative displacement parallel $D_{||}$ and perpendicular $D_{\perp}$ to the worm orientation in sediments (E) and water (F) over $T = 40$\,s.  
} 
\label{fig:waves}
\end{figure}

The projected shapes of a worm of length $l_w = 26.6$\,mm and $d_w = 0.5$\,mm moving in the sediments are shown in Fig.~\ref{fig:waves}A as it travels approximately its body length. We observe that it traces a narrow path in the medium, similar in form to that observed in the sedimented bed in the $W_c = 12.7$\,mm wide container shown in Fig.~\ref{fig:img}C. To compare and contrast the observed burrowing dynamics with swimming, we show the worm shapes recorded when constrained between two parallel plates separated by 2\,mm and immersed in container filled with water in Fig.~\ref{fig:waves}B. (A schematic can be found in SI Appendix, Fig.~S4A.) Because the worm does not swim up very high above the surface unless threatened, we measure the motion when the quasi-2D container is horizontal so that the strokes and the trajectories can be compared while being constrained similarly. Over the same time window, we observe that the worm moves forward only a fraction of its body length while performing somewhat larger undulatory strokes in water. If sediments are also added in this horizontal orientation of the observation plane, corresponding to near-zero overburden pressure as near the surface (see SI Appendix, Effect of Sediment Consolidation on Anchoring), we observe the path is narrower compared to that in water, but not as narrow as in the vertical orientation shown in Fig.~\ref{fig:waves}, where the weight of the grains above pushes on the worm  to constrain it to move in a still tighter path (see SI Appendix, Fig.~S4B).

To compare the form of the strokes used while burrowing and swimming, we use the body reference frame which is oriented along the average direction in which the worm points, and where its origin is located at the worm centroid as shown in Fig.~\ref{fig:img}B. We plot the same measured shapes in Fig.~\ref{fig:waves}C and D in the body reference frame, but where the centroid is shifted vertically by time denoted by the vertical axis. We observe that transverse undulations and and that the worm elongations and length contraction can be observed in fact in both mediums. To show that the shortening of the worm indeed arises due its changes, and not simply because of the transverse undulations, we obtain the difference of dynamic worm length from its mean length $\Delta l_w(t)$ and use the color bar to render each snapshot. Periodic changes in the length can be unambiguously observed.   

\begin{figure}
\begin{center}
\includegraphics[width=0.45\textwidth]{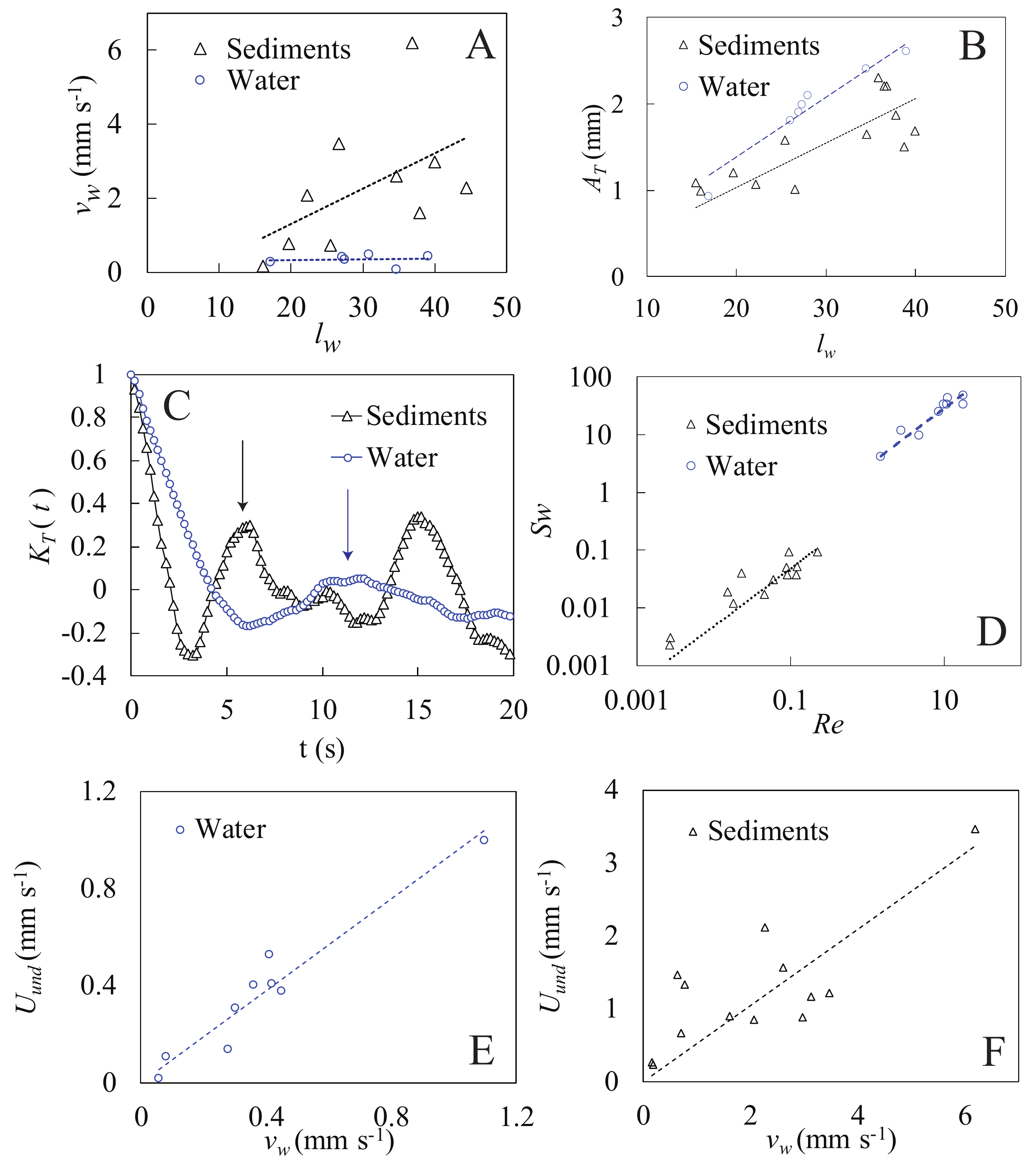}
\end{center}
\caption{(A) The measured worm speed versus its length shows significant scatter from worm to worm. However, $v_w$ is overall higher in the sediments compared with water. (B) The transverse amplitude $A_T$ increases approximately linearly with $l_w$. (C) $K_T(t)$ shows a peak at $T_T$ (indicated by arrow) corresponding to the transverse undulations time scale. (D) $Sw$ versus $Re$ is observed to scale roughly linearly. (E-F)  Calculated $U_{und}$ versus $v_w$ using resistive force theory assuming low amplitude and low-$Re$ conditions in water (E) and the sediments (F). The line slope is 0.94 with goodness of fit R$^2 = 0.93$ in water, and a slope 0.52 and R$^2 = 0.49$ in the sediments (see text). Peristaltic motion, besides undulation, contributes to $v_w$ leading to the lower correlation in the sediments.} 
\label{fig:trans}
\end{figure}

Then, we obtain the component of worm speed over a short time interval $\Delta t$ parallel to the average body axis $v_{||}  = \Delta {\bf R}_c \cdot \hat{x}/\Delta t$, and the speed perpendicular to the average body axis $v_{\perp} = \Delta {\bf R}_c \cdot \hat{y}/\Delta t$. The resulting cumulative displacement of the worm in the direction parallel to the average body axis $D_{||}(t) = \sum_0^t v_{||} \Delta t$, and perpendicular to the average body axis $D_{\perp} = \sum_0^t v_{\perp} \Delta t$ are plotted in Fig.~\ref{fig:waves}E,F over a longer observation time $T =40$\,s, with $\Delta t = 200$\,ms.  We observe that the worm moves on average along the direction of the body orientation in both mediums. Hence, the average forward speed of the worm $v_w \approx D_{||}/T$. 
Fig.~\ref{fig:trans}A shows the worm locomotion speed $v_w$ measured over a time interval $T \geq 20$\,s as a function of $l_w$ using 10 different worms in the sediments, and 6 different worms in water listed in SI Appendix, Table~S1. Besides higher speeds in the sediments, an overall increasing trend with $l_w$ is found with significant variations from worm to worm due to behavioral differences. In the following, we focus on understanding the measured speed in terms of the strokes used by the worm and the rheology of the medium.

\subsection*{Transverse Body Undulations} 
We measure the transverse undulations using the root mean square (RMS) transverse amplitude in the body frame of reference $A_T = \sqrt{\langle y^2 \rangle}$, where $\langle .. \rangle$ denotes averaging over the length of the worm. 
Fig.~\ref{fig:trans}B shows a plot of $A_T$ where we observe that it increases roughly linearly with $l_w$, and with slightly higher slope in water compared with the sediments ($A_T/l_w = 0.069$ and goodness of linear regression R$^2 =0.94$ versus $A_T/l_w = 0.052$ and R$^2 =0.67$). Further, by characterizing the worm orientation correlation function as discussed in SI Appendix, Worm-Body Orientation Correlations, we find its persistence length to be of the order of its length in both mediums. Thus, the worm can be considered as rod-like with undulation amplitude $A_T/l_w \ll 1$ in both mediums.  

Then, to determine the relevant regime for its dynamics, we use the Reynolds number $Re = \rho v_w l_w/\mu$, where $\rho$ and $\mu$ are the density and viscosity of the medium, respectively, and the Swimming number $Sw = \rho v_T l_w/\mu$, where $v_T$ is the velocity associated with the transverse oscillations which determines propulsion~\cite{Gazzola2014}. For two-dimensional sinusoidal oscillations with amplitude $B$ and frequency $f_T$, $v_{T} = 2 \pi f_T B$, and  we have with $B=\sqrt{2} A_T$, $Sw  = 2\sqrt{2} \pi f_T A_T l_w/\mu$. To estimate $f_T$, we use the displacement $y_c(t)$ corresponding to the mid-section of the worm in the body frame of reference, and calculate the transverse amplitude correlation function $K_T(t) = \langle y_c(t_o + t) y_c(t_o) \rangle / \langle y_c^{2} \rangle$ averaged over time $t_o$. From Fig.~\ref{fig:trans}C, we observe that $K_T(t)$ oscillates and shows peaks which correspond to the period of transverse oscillations. We use the time at which the first strong peak occurs with the time scale $T_T$ and determine $f_T = 1/T_T$. Then, we find from the plot of $Sw$ and $Re$ in Fig.~\ref{fig:trans}D that the estimated $Re$ ranges between 1 and 20 in the case of water,  corresponding to the crossover regime where viscous and inertia effects can be important. Nonetheless, we find that $Sw$ increases linearly with $Re$ consistent with what may be expected by resistive force theory in the low-$Re$ and low-amplitude regime~\cite{lighthill76}, where the swimming speed of an undulating filament   
\begin{equation}
U_{und} = \frac{2 \pi^2 f_T  B^2}{\lambda} (\xi_r -1),
\label{eq:und}
\end{equation}
where $\xi_r$ is the ratio of the drag in the perpendicular and parallel direction w.r.t. the rod axis. We plot this estimated swimming speed versus the measured speed in Fig.~\ref{fig:trans}E using $\lambda \approx l_w$  and $\xi_r = 2$ for water~\cite{cox70}, and find excellent agreement. Then, multiplying  Eq.~\ref{eq:und} by $\rho l_w/\mu$ on both sides, we have 
\begin{equation}
Sw = \frac{l_w}{\sqrt{2}\pi A_T(\xi_r-1)} Re,
\label{eq:Sw}
\end{equation}
which corresponds to the line drawn in Fig.~\ref{fig:trans}D using $A_T/l_w = 0.069$ and $\xi_r =2$.  

\subsection*{Drag-assisted Propulsion in Sediments} 
To find the appropriate drag encountered by the worm while undulating in the sediments,  we performed complementary measurements with a thin rod corresponding to the worm body over the typical range of speeds encountered by the worm. As discussed in more detail in SI Appendix, Drag Measurements, we observe a drag which increases sublinearly with speed and depends on the orientation of the rod axis relative to the direction of motion. Over the range of speeds $U$ from $0.1$ to $10$\,mm\,s$^{-1}$ relevant to the motion of the worm, we measure the effective viscosity of the medium $\eta_{e} \approx 4.0 \times U^{-0.63}$\,Pa\,s, and $\xi_r \approx 6$ as shown in SI Appendix, Fig.~S6. This measured $\xi_r$ is similar in magnitude, but somewhat lower than the drag anisotropy of approximately 10 reported in dry granular matter~\cite{maladen09}.  

The estimated $Re$ and $Sw$ corresponding to the worm motion in the sediments are also shown in Fig.~\ref{fig:trans}D. They are systematically lower than that for water because the effective viscosity in the medium is essentially a thousand times higher than water, while the worm parameters remain essentially unchanged. We also observe that $Sw$ versus $Re$ can be described by a linear fit. Now, it has been shown that resistive force theory may be applied to sandfish lizards moving through dry sand~\cite{maladen09}. 
Thus, notwithstanding the shear thinning nature of the medium, we may expect Eq.~\ref{eq:und} to capture the undulatory component of the worm speed. We plot the estimated $U_{und}$ versus $v_w$ with $\xi_r = 6$ for sediments in Fig.~\ref{fig:trans}D and observe that the data can be described by a linear fit, but with a slope of $0.52$ and R$^2 = 0.49$ indicating the presence of other factors which contribute to the observed $v_w$. Now, if shear thinning effects of the medium were important, it would lead to a lower estimate of $U_{und}$~\cite{Riley2017} which is opposite to the trend needed to capture the lower slope. Thus, to understand the difference, we examine next the contribution of the observed peristaltic motion to the locomotion speed of the worm.  

\subsection*{Peristaltic Motion} 
\begin{figure}
\begin{center}
\includegraphics[width=0.45\textwidth]{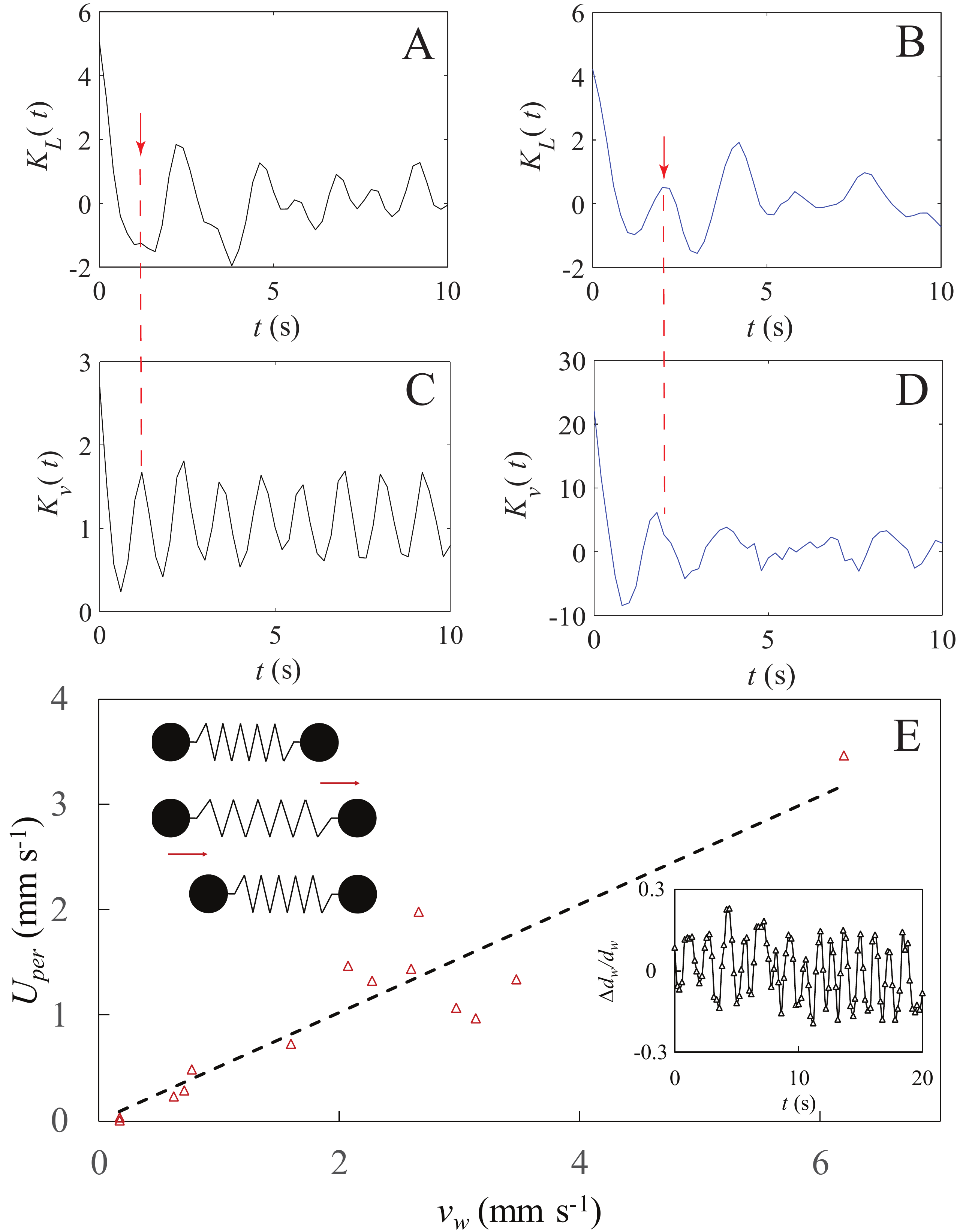}
\end{center}
\caption{(A,B) The length correlation function $K_L(t)$  in sediments (A) and in water (B). (C,D) The velocity correlation $K_v(t)$  in sediments (C) and in water (D). The peaks are correlated corresponding to peristaltic motion in the sediments. (E) The measured worm speed $v_w$ versus the calculated peristaltic speed $v_{per}$ using the anchor model. The dashed line corresponds to a linear fit with slope $0.5 \pm 0.1$. Insets: Illustration of the anchor model, and  the fractional change in worm diameter at its center over time.              
} 
\label{fig:peris}
\end{figure}

To extract the dominant oscillation frequency, we obtain the longitudinal amplitude correlation function $K_L(t) = \langle \Delta l_w(t_o + t) \Delta l_w(t_o) \rangle / \langle \Delta l_w^2 \rangle$, where $\langle .. \rangle$ denotes averaging over time $t_o$, and $\langle \Delta l_w^2 \rangle$ is the mean square fluctuation over $T$. Fig.~\ref{fig:peris}A and in Fig.~\ref{fig:peris}B shows plots of $K_L(t)$ in case of the sediments and water, respectively. Peaks corresponding to the dominant periods can be observed in both mediums. We associate the longitudinal oscillation period $T_L$ with the first peak, which is observed to be stronger and clearer in the sediment example. 

Then, we calculate the velocity correlation function $K_v(t) = \langle v_{w}(t_o + t) v_{w}(t_o) \rangle/v_w^2$, which measures the correlation of $v_{w}$ at time $t_o$ with that after a short time $t$. Because of the normalization by $v_w^2$, $K_v(t)$ can be expected to oscillate or approach 1 over time. $K_v(t)$ is plotted in Fig.~\ref{fig:peris}C and Fig.~\ref{fig:peris}D in the sediments and water, respectively, and show oscillations about the average value in both mediums. In the sediments, $K_v(t)$ always remains positive while oscillating and remains strong over the duration plotted.  Whereas, $K_v(t)$ becomes negative in water, indicating a back and forth motion at short time scales, before decaying rapidly to the mean value. Further, comparing the peaks in $K_L(t)$ with those in $K_v(t)$ in the sediments, we observe that the dominant oscillation period $T_L$ in $K_L(t)$ is twice the period $T_v$ in $K_v(t)$.  In comparison, the first peak in $K_L(t)$ and $K_v(t)$ appear at the same time in water. 

\subsection*{Anchor Model}
To understand these correlations, we consider an idealized anchor model of peristaltic motion~\cite{tanaka12} as illustrated in the inset to Fig.~\ref{fig:peris}E. Here, the worm is represented in the form of a pair of beads connected by a spring which travels forward by elongating its body through a length $\epsilon$ while anchoring its tail, and then moving forward by contracting its body through the same length $\epsilon$ while anchoring its head to regain its initial length, as indicated by the arrows. Then, the distance moved by the worm centroid is $\epsilon/2$ during the expansion as well as the contraction phase, and the net displacement is $\epsilon$ in each period of oscillation. On the other hand, if the worm is not anchored, but slips similarly during  expansion and contraction phase, then the net displacement can be expected to be less than $\epsilon$ by a factor $\alpha$, which is between 0 and 1. 

Assuming that the primary oscillation of the worm length occurs sinusoidally with period of longitudinal oscillation $T_L$, we have $\Delta l_w (t) = A_L  \sin{(2\pi t/T_l)} + \Delta l_s(t)$, where $\Delta l_s(t)$ captures additional time dependence with $\langle \Delta l_s(t) \rangle =0$. Then, $\langle \Delta l_w(t_o) \Delta l_w(t_o+t) \rangle= \frac{1}{2}{A_L^2} \cos{(2\pi t/T_l)} +  \langle \Delta l_s^2 \rangle$. If $\Delta l_s(t)$ decays rapidly compared with $T_l$, then the peaks associated with the sinusoidal mode in $K_l(t)$ can be expected have constant amplitude $A_L^2/2$ after the initial rapid decay as is seen in Fig.~\ref{fig:peris}A. Because $\epsilon = 2 A_L$ and $T_v = T_L/2$, 
we can then estimate the speed due to peristaltic motion to be  $U(t) = U_{per}(1 + \cos{4\pi t/T_L})$ to leading order, with 
\begin{equation}
U_{per} = 2 \alpha  A_L/T_L.
\end{equation} 
Obtaining $A_L$ from the strength of the first peak in $K_L(t)$, we plot the calculated speed $U_{per}$ versus the measured speed $v_w$ in Fig.~\ref{fig:peris}E. We find that $U_{per}$ versus $v_w$ can be fitted by a  straight line with a slope $U_{per}/v_w = 0.5 \pm 0.1$. In arriving at these values, we have ignored other oscillation frequencies which may contribute to the peristaltic motion. An upper bound  $U_{per}/v_w = 0.85$  can be estimated using $K_L(t)$ corresponding to $t =0$\,s which includes all length fluctuations. Some of these contributions may offset the fact that worm does not travel in a straight line, but rather in a sinusoidal path.  Hence, we conclude that the peristaltic motion contributes at least equally to the locomotion speed of these worms in the sediments.

\begin{figure}
\begin{center}
\includegraphics[width=0.3\textwidth]{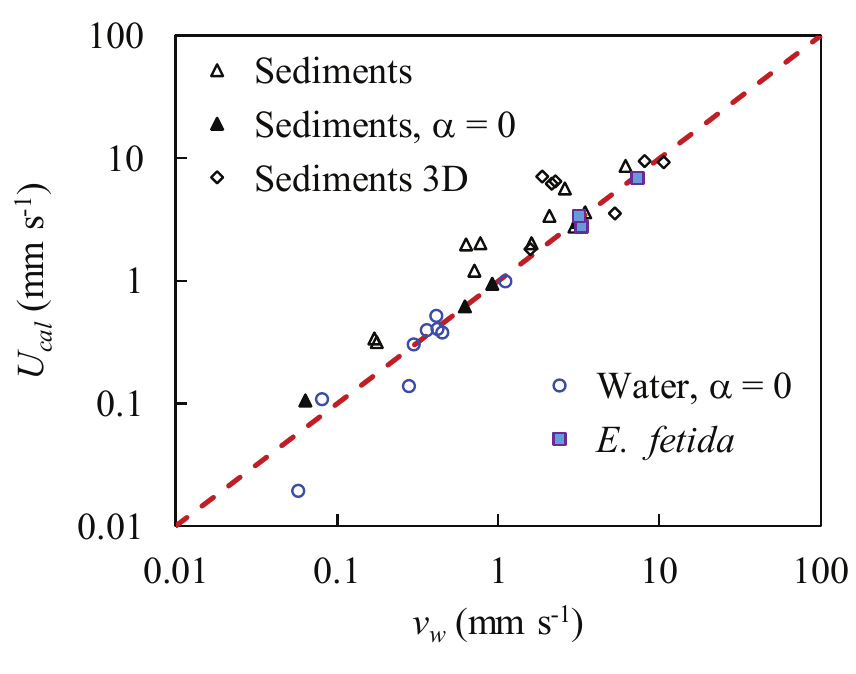}
\end{center}
\caption{The calculated versus the measured speeds corresponding to the trails listed in SI Appendix, Table~S1 are in good agreement. A dashed line with slope 1 is drawn for reference.               
} 
\label{fig:vel}
\end{figure}

\begin{figure*}
\begin{center}
\includegraphics[width=0.8\textwidth]{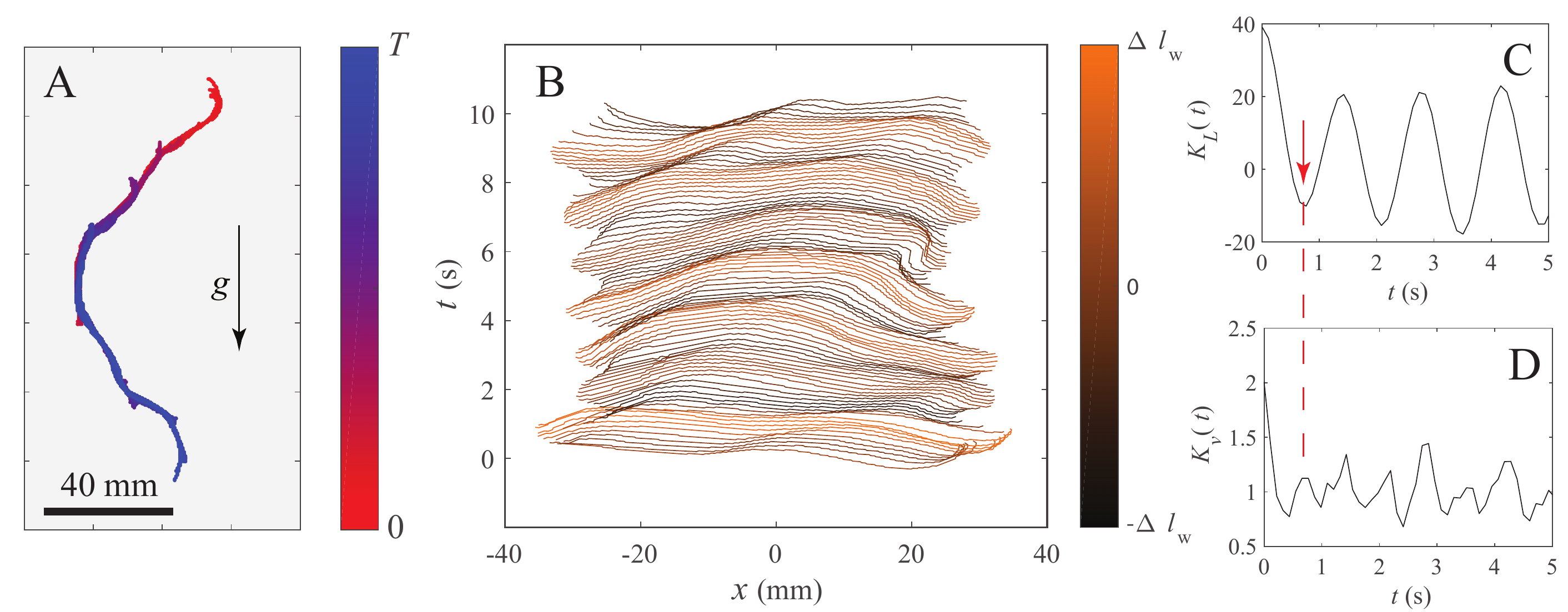}
\end{center}
\caption{Tracked snapshots of the composting earthworm {\textit E. fetida} wof length $l_w = 68.6$\,mm over $T = 10$\,s at $\Delta t = 109$\,ms time intervals in the lab reference frame (A) and body reference frame (B) as it burrows through water-saturated granular sediments. The progress of time is denoted by colors according to the color bar in (A), and the length relative to the mean length is denoted according to the color bar in (B).  {\it E. fetida} shows peristaltic motion and transverse undulations similar to those observed in {\it L. variegatus} in Fig.~\ref{fig:waves}A. The peaks in the length correlation function (C) occurs at approximately at twice the time interval compared to those in the velocity correlations (D).              
} 
\label{fig:fetida}
\end{figure*}
 
\subsection*{Medium Rheology and Locomotion Speed}
{\it L. variegatus} is known to dynamically deploy 10-20 micron long hair-like projections called chaetae~\cite{marchese15} along with muscle contractions to change the relative friction during the sliding and anchoring phase while crawling on solid surfaces~\cite{drewes99a}. Given the small size of chaetae relative to $d_w$, deploying them has negligible effect on the drag experienced by a worm in a fluid which obeys non-slip boundary conditions. Thus, the anchoring parameter $\alpha$ can be expected to be approximately zero in water and other Newtonian fluids. By contrast, the drag experienced by an intruder in sediments can depend sensitively on the normal force acting between the grains in the medium and its surface~\cite{panaitescu17,jewel18}. Even if the friction between the worm and the grains can be changed, anchoring cannot be achieved if the grains are very loosely packed as when the overburden pressure is nearly zero or when grains simply move out of way as near the bed surface as discussed in SI Appendix, Effect of Sediment Consolidation on Anchoring. 

Thus, the yield-stress nature of the medium is also important to achieving  anchoring needed for peristaltic motion. This strength characterize by $\mu_o$ in turn depends on the volume fraction of the sediments $\phi$. If $\phi$ is below a critical value $\phi_c$, the medium can flow and $\mu_o$ can be expected to vanish.  When $\phi \rightarrow \phi_c \approx 0.6$, corresponding to the volume fraction of the sedimented bed~\cite{panaitescu17}, the granular medium jams and the yield strength increases rapidly within a few percent of this value. Because we observe peristaltic motion, we conclude that the worm can dynamically anchor itself by manipulating $\phi$ near its body by changing its diameter as shown in the Inset to Fig~\ref{fig:peris}E, in addition deploying chaetae to vary the friction between its body and the medium. 

Then, assuming that the locomotion speed of the worm occurs due to a linear superposition of the contributions due to the undulatory and the peristaltic motion, we have $U_{cal} = U_{und} + U_{per}$. Fig.~\ref{fig:vel} shows a plot of $U_{cal}$ versus $v_w$ in the sediments and in water corresponding the trails listed in SI Appendix, Table~S1. In case of water and shallow sediment beds, as discussed $\alpha = 0$ due to absence of anchoring, and only the contribution of the undulatory stroke is included. We observe that all the data is clustered around the slope 1 line showing good agreement between the calculated and measured speeds in both mediums. Thus, we find that the burrowing speed of {\it L. variegatus} in water-saturated sediments is determined by a combination of drag-assisted propulsion provided by the transverse undulatory motion, and peristaltic motion along its body. Whereas, only the transverse undulatory motion is important to its swimming speed in water and near the bed surface where the overburden pressure goes to zero. 

\subsection*{Dual Strokes in  \textbf{\textit{ E. fetida}}}
We further investigate the dynamics of the common composting earthworm {\it Eisenia fetida} to examine if dual peristaltic and undulatory strokes are observed in other organisms as well. The general behavior of the earthworms and their physical characteristics are discussed in SI Appendix, Earthworm Dynamics.  Figure~\ref{fig:fetida}A shows the projected shapes of the earthworm as it 
burrows straight down in a container with $L_c = 50$\,cm, $H_c = 30$\,cm and $W_c = 28$\,mm filled with the same sediment medium over $T = 10$\,s. As in the case of {\it L. variegatus} shown in Fig.~\ref{fig:waves}A, we observe that the earthworm burrows in a narrow sinusoidal path which is not much wider than its body except near its head and tail. Then, by plotting the same snapshots as for {\it L. variegatus} in the body reference frame in Fig.~\ref{fig:waves}B, we observe that {\it E. fetida} also performs undulatory strokes as well as elongation-contraction strokes. 

To access the period of the longitudinal stroke, we plot $K_L(t)$ in Fig.~\ref{fig:fetida}C, and observe a clear peak at $T_L = 1.6$\,s. By comparison, a peak in $K_v(t)$  is observed at approximately half of $T_L$ corresponding to the signature of peristaltic motion which was also seen {\it L.variegatus} as in Fig.~\ref{fig:peris}C and E. Accordingly,  we calculate $U_{per} = 4.5$\,mm\,s$^{-1}$ using Eq.~\ref{eq:und} over a time interval where its average speed is measured to be 7.3\,mm\,s$^{-1}$, or approximately 61\%, assuming $\alpha =1$. Then, we obtain $B$ from the transverse undulations to estimate $U_{und}$. We estimate $T_T = 15.4 \pm 1$\,s from the peak in $K_T$ shown in SI Appendix, Fig.~S7C. Then, we obtain $U_{und} = 2.4$\,mm\,s$^{-1}$ or approximately 33\% of the measured speed. Thus, a combination of peristaltic and undulatory strokes contribute to the observed burrowing speed of {\it E. fetida} in water-saturated sediments as well. This data, along with two other data sets have been also added to Fig.~\ref{fig:vel}, and observed to be in overall agreement with the calculated values. 

\section*{Discussion and Concluding Remarks}
Thus, by directly tracking {\it L. variegatus} and {\it E. fetida} inside water and sediments, we have demonstrated that limbless worms can move in mediums with wide ranging rheological properties using a combination of peristaltic and undulatory strokes. While the stroke amplitude can be modified somewhat as evidenced by the decrease of transverse undulations in sediments compared to in water, the nature of the medium has a significant impact on the effectiveness of these strokes. When the worm cannot anchor itself, as when moving in water or near the surface of a sedimented bed, our study shows that only the transverse undulatory motion is important to achieving locomotion. But deep in the sediment bed, where the overburden pressure causes the grains to stay in close contact, we observe clear importance of peristaltic motion in achieving locomotion. Conversely, lacking the dual strokes, a worm would be at a disadvantage while swimming in the water or in the unconsolidated grains very near the bed surface. Whereas, an undulating worm would be increasingly at a disadvantage as it burrows deeper because the drag in moving perpendicular to its body grows more rapidly than parallel to its body making that motion prohibitive at large enough depth. In fact, peristaltic motion can be expected to dominate as an active intruder moves deeper based on our analysis. This suggests that active intruders, whether biological or synthetic, can be designed to exploit these dual strokes to move between fluid-like regions with negligible yield stress and frictional granular regions with large yield stress.

\section*{Methods}
\subsection*{Specimens}
 {\it L. variegatus} were obtained from Carolina Biological Supply Company (https://www.carolina.com) on October 3, 2017, and were sustained in a fresh water aquarium. {\it E. fetida} were obtained from Uncle Jim's Worm Farm (https://unclejimswormfarm.com/) on June 6, 2017, and were maintained in a wet soil filled container. The worms used to perform quantitative measurements are listed in SI Appendix, Table~S1. Both were housed in an air conditioned lab maintained at $24 \pm 2\,^\circ$C.%

\subsection*{Medium} 
We use clear hydrogel grains (Acrylic Acid Polymer Sodium Salt, Sumitomo Seika Chemicals Co.) with density $\rho_g =1.004$\,g\,cm$^{-3}$ and diameter $d_g$ ranging from 0.5\,mm to 2\,mm that are fully immersed in water as the sediment medium. Its volume fraction in the sedimented bed is measured to be $\phi_g = 0.6 \pm 0.01$. 


\subsection*{Worm tracking} Because the refractive indexes of these grains and water are essentially the same, the medium appears transparent, allowing us to visualize the worm dynamics through transparent glass sidewalls. Images are thresholded to identify a connected set of pixels associated with the worm body. Its head, tail, and skeletal shape are then found using the operation {\it bwmorph} in the Image Processing Toobox in MATLAB.




\subsubsection*{SI Datasets} 
The data corresponding to the measured worm speed, worm length, specimens and containers used can be found in the SI Appendix and in SI Datasets S1-S5.  




\showmatmethods{} 

\acknow{We thank Andreea Panaitescu, Gregory Jones, Rausan Jewel, and Benjamin Allen  for assistance with experiments, and Alex Petroff for comments on the manuscript. This work was supported by the National Science Foundation under grant number CBET-1805398. }

\showacknow{} 


\begin{thebibliography}{10}

\bibitem{hosoi15}
Hosoi AE, Goldman DI (2015) Beneath our feet: Strategies for locomotion in
  granular media.
\newblock {\em Annual Review of Fluid Mechanics} 47:431--453.

\bibitem{quillin99}
Quillin KJ (1999) Kinematic scaling of locomotion by hydrostatic animals:
  ontogeny of peristaltic crawling by the earthworm lumbricus terrestris.
\newblock {\em Journal of Experimental Biology} 202:661--674.

\bibitem{tanaka12}
Tanaka Y, Ito K, Nakagaki T, Kobayashi R (2012) Mechanics of peristaltic
  locomotion and role of anchoring.
\newblock {\em Proceedings of the Royal Society Interface} 9:222--233.

\bibitem{gidmark11}
Gidmark NJ, Strother JA, Horton JM, Summers AP, Brainerd EL (2011) Locomotory
  transition from water to sand and its effects on undulatory kinematics in
  sand lances (ammodytidae).
\newblock {\em Journal of Experimental Biology} 214:657--664.

\bibitem{dorgan13}
Dorgan KM, Law CJ, Rouse GW (2013) Meandering worms: mechanics of undulatory
  burrowing in muds.
\newblock {\em Proceedings of the Royal Society B} 280:20122948.

\bibitem{dorgan15}
Dorgan KM (2015) The biomechanics of burrowing and boring.
\newblock {\em Journal of Experimental Biology} 218:176--183.

\bibitem{fang-yen2010}
Fang-Yen C, et~al. (2010) Biomechanical analysis of gait adaptation in the
  nematode caenorhabditis elegans.
\newblock {\em PNAS} 107:20323–20328.

\bibitem{jung10}
Jung S (2010) {\it Caenorhabditis elegans} swimming in a saturated particulate
  system.
\newblock {\em Physics of Fluids} 22:031903.

\bibitem{lockery2008}
Lockery SR, et~al. (2008) Artificial dirt: Microfluidic substrates for nematode
  neurobiology and behavior.
\newblock {\em J. Neurophysiol.} 99:3136--3143.

\bibitem{gray53}
Gray J (1953) Undulatory propulsion.
\newblock {\em Journal of Cell Science} s3-94:551--578.

\bibitem{lighthill76}
Lighthill J (1976) Flagellar hydrodynamics.
\newblock {\em SIAM Review} 18:161--230.

\bibitem{lauga09}
Lauga E, Powers TR (2009) The hydrodynamics of swimming microorganisms.
\newblock {\em Rep. Prog. Phys} 72:096601.

\bibitem{panaitescu17}
Panaitescu A, Clotet X, Kudrolli A (2017) Drag law for an intruder in granular
  sediments.
\newblock {\em Physical Review E} 95:032901.

\bibitem{jewel18}
Jewel R, Panaitescu A, Kudrolli A (2018) Micromechanics of intruder motion in
  wet granular medium.
\newblock {\em Physical Review Fluids} 3:084303.

\bibitem{allen19}
Allen B, Kudrolli A (2019) Effective drag of a rod in fluid-saturated granular
  beds.
\newblock {\em Phys. Rev. E} 100:022901.

\bibitem{balmforth14}
Balmforth NJ, Frigaard IA, Ovarlez G (2014) Yielding to stress: Recent
  developments in viscoplastic fluid mechanics.
\newblock {\em Annual Review of Fluid Mechanics} 46:121--146.

\bibitem{drewes99a}
Drewes C, Cain K (1999) As the worm turns: Locomotion in a freshwater
  oligochaete worm.
\newblock {\em The American Biology Teacher} 46:438--442.

\bibitem{Gazzola2014}
Gazzola M, Argentina M, Mahadevan L (2014) Scaling macroscopic aquatic
  locomotion.
\newblock {\em Nature Physics} 10:758.

\bibitem{cox70}
Cox RG (1970) The motion of long slender bodies in a viscous fluid part 1.
  general theory.
\newblock {\em Journal of Fluid Mechanics} 44(4):791–810.

\bibitem{maladen09}
Maladen RD, Ding Y, Li C, Goldman DI (2009) Undulatory swimming in sand:
  subsurface locomotion of the sandfish lizard.
\newblock {\em Science} 325:314--318.

\bibitem{Riley2017}
Riley EE, Lauga E (2017) Empirical resistive-force theory for slender
  biological filaments in shear-thinning fluids.
\newblock {\em Physical Review E} 95:062416.

\bibitem{marchese15}
Marchese MR, dos Santos MR, dos Santos~Lima JC, Pamplin PAZ (2015) First record
  of introduced species lumbriculus variegatus müller, 1774 (lumbriculidae,
  clitellata) in brazil.
\newblock {\em BioInvasions Records} 4:81--85.

\end{thebibliography}

\end{document}